\documentclass[sigconf]{acmart}
\AtBeginDocument{%
  \providecommand\BibTeX{{%
    \normalfont B\kern-0.5em{\scshape i\kern-0.25em b}\kern-0.8em\TeX}}}

\usepackage[english]{babel}
\usepackage{blindtext}
\usepackage{graphicx}
\usepackage[breakable, skins]{tcolorbox}
\usepackage{csquotes}
\usepackage{paralist}
\usepackage{subcaption}
\usepackage{graphicx}
\usepackage{color, soul}
\usepackage{setspace}
\usepackage[toc, page]{appendix}
\usepackage{cleveref}
\usepackage{multirow}
\usepackage{varwidth}
\usepackage{dblfloatfix}
\usepackage{balance}
\usepackage{chngcntr}
\usepackage{epstopdf}
\usepackage{layouts}
\usepackage{microtype}
\usepackage{gensymb}
\usepackage{adjustbox}
\usepackage{ifthen}
\usepackage{url}

\usepackage{breakurl}

\copyrightyear{2024} 
\acmYear{2024} 
\setcopyright{rightsretained} 
\acmConference[GE@ICSE '24]{2024 ACM/IEEE Workshop on Gender Equality, Diversity, and Inclusion in Software Engineering}{April 20, 2024}{Lisbon, Portugal}
\acmBooktitle{2024 ACM/IEEE Workshop on Gender Equality, Diversity, and Inclusion in Software Engineering (GE@ICSE '24), April 20, 2024, Lisbon, Portugal}\acmDOI{10.1145/3643785.3648494}
\acmISBN{979-8-4007-0575-5/24/04}

\begin{document}

\author{Sonja M.\ Hyrynsalmi}
\email{sonja.hyrynsalmi@lut.fi}
\orcid{0000-0002-1715-6250}
\affiliation{%
  \institution{LUT University}
  \city{LUT}
  \country{Finland}
}

\author{Ella Peltonen}
\email{ella.peltonen@oulu.fi}
\orcid{0000-0002-3374-671X}
\affiliation{%
  \institution{University of Oulu}
  \city{Oulu}
  \country{Finland}
}

\author{Fanny Vainionp\"a\"a}
\email{fanny.vainionpaa@oulu.fi}
\affiliation{%
  \institution{University of Oulu}
  \city{Oulu}
  \country{Finland}
}

\author{Sami Hyrynsalmi}
\email{sami.hyrynsalmi@lut.fi}
\orcid{0000-0002-5073-3750}
\affiliation{%
  \institution{LUT University}
  \city{LUT}
  \country{Finland}
}

\renewcommand{\shortauthors}{Hyrynsalmi et al.}

\title{The Second Round: Diverse Paths Towards Software Engineering}

\begin{abstract}
  In the extant literature, there has been discussion on the drivers and motivations of minorities to enter the software industry. For example, universities have invested in more diverse imagery for years to attract a more diverse pool of students. However, in our research, we consider whether we understand why students choose their current major and how they did in the beginning decided to apply to study software engineering. We were also interested in learning if there could be some signs that would help us in marketing to get more women into tech. We approached the topic via an online survey (N = 78) sent to the university students of software engineering in Finland. Our results show that, on average, women apply later to software engineering studies than men, with statistically significant differences between genders. Additionally, we found that marketing actions have different impacts based on gender: personal guidance in live events or platforms is most influential for women, whereas teachers and social media have a more significant impact on men. The results also indicate two main paths into the field: the traditional linear educational pathway and the adult career change pathway, each significantly varying by gender.
\end{abstract}


\keywords{Software Engineering Education, Gender, University Student Admission}

\maketitle

\section{Introduction}

Research has shown that the primary factors influencing students' choice of study program in information systems are career-oriented. The most significant factor for students when selecting a study program was identified as job availability, closely followed by job security, career opportunities, and the prospect of engaging in work assignments.~\cite{acilar2022exploring} It has also been highlighted that students in Information and Communication Technology (ICT) generally share similar perceptions of the field with students from other areas, with the notable exception that they view ICT as more creative and people-oriented than students in the other fields. This insight underscores the importance of highlighting the creative aspects of software engineering to potentially attract a broader and more diverse student body, including increasing the number of women in ICT studies.~\cite{rajala2022perceiving}. 

\begin{figure}
    \centering
    \includegraphics[width=0.9\linewidth]{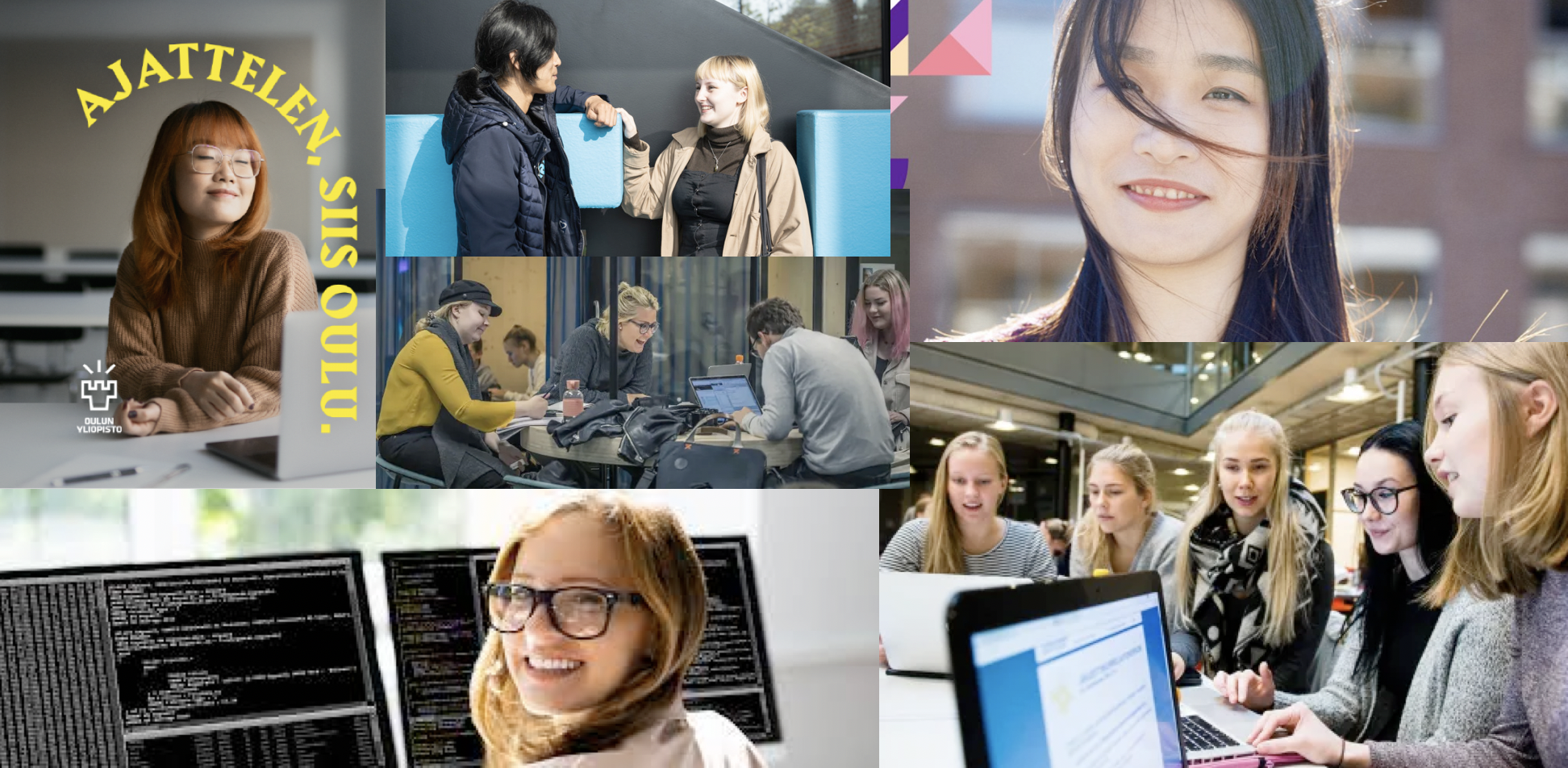}
    \caption{Some example pictures used to market software engineering (or ICT field as a whole) for BSc degree studies. Pictures are quoted from the Universities' online materials and StudyInfo, the governmental application website.}
    \label{fig:kollaasi}
\end{figure}

However, although more attention has been paid to the more diverse marketing, for example, the images used in the university websites and marketing (Figure~\ref{fig:kollaasi}), the number of women in software engineering studies has increased slowly. This also affects the representation of women in software engineering, as fewer women are transitioning to working life~\cite{hill2010so}. Women face different socio-cultural challenges (such as work-life balance issues, impostor syndrome and a lack of recognition and peer equality) in their studies and working life~\cite{margolis2002unlocking, trinkenreich2022empirical} and attracting women to the software engineering field is that way harder, although women would have the same skills as men.

To get a more timely understanding of the gender differences between different marketing strategies in software engineering education, we conducted an online survey (n = 78) for university-level software engineering students in Finland. In this research paper, we provide preliminary insights into our data, which will be extended with further interviews and more detailed questionnaires about reasons behind applying to software engineering programs and marketing efforts that may have impacted the applicants' decisions. With a preliminary answer pool consisting of half-female and half-male attendants, we can present the first comparison efforts in this paper.

This study addresses the following research questions (RQ):
\begin{description}
    \item[RQ1] What are the perceived reasons for selecting software engineering university studies, and how do they differ between genders?
    \item[RQ2] When, where, and how do people of different genders find information and influence the decision to apply for software engineering degrees?
\end{description}



The remaining sections are structured as follows: Section~\ref{sec:back} reviews related work and central concepts. Section~\ref{sec:research} presents the empirical research approach, and the results are shown in Section~\ref{sec:results}. Discussion and conclusions are given in Sections~\ref{sec:disc} and~\ref{sec:conc}.

\section{Background}\label{sec:back}

\subsection{Women studying STEM} 

Why don't more females choose the technology field? The question is discussed in research publications and in many university marketing meetings. The reasons have been found in stereotypical thinking of female and male career paths, which can already manifest in a lack of female teachers in the mathematics and natural sciences~\cite{bottia2015growing,canaan2023impact}. Another possible reason is seen as a non-technical self-assessment of female students, guided by expectations of society and families~\cite{ertl2017impact,fisher2020gender,tahsin2022can}. However, when they complete their studies, female graduates have little to no difference in skills from their male counterparts~\cite{vooren2022comparing}, underlining how important it is to get that skilled workforce interested in the technology industry. 

Successful marketing to gather more female students in the STEM (Science, Technology, Engineering, and Mathematics) field is often seen as a representation problem. Solutions include increasing the number of females in images used in student marketing campaigns and publishing stories featuring female technology professionals as role models~\cite{milgram2011recruit, hyrynsalmi2019software}. Other suggestions include organising more hands-on events, such as all-female coding workshops and tech seminars for non-technology-oriented students. Indeed, such hands-on activities seem to arouse interest towards the STEM career~\cite{ertl2017impact,donmez2021impact, hyrynsalmi2019underrepresentation}. Equally, access to a "real" female STEM professional increases female students' self-esteem regarding their potential and self-perceived skills towards the STEM career~\cite{gonzalez2020girls,kelly2019stem}. 

\subsection{Software Engineering in diversity crisis} 

Diversity in software engineering has explicitly called to be in a crisis~\cite{albusays2021diversity}. Gender-diverse teams have been reported to perform better in software engineering projects~\cite{russo2020gender,catolino2019gender}. Thus, getting more females present in the field could improve the quality of the results and the knowledge-sharing culture as a whole~\cite{zolduoarrati2021value} and boost innovation and creativity activities~\cite{kohl2022benefits}. There is a common belief that women would pursue IT because of passion and men because of material concerns\cite{soylu2020cultural}. However, some studies also show that female and male professionals value actually similar features in their software engineering jobs: good work-family balance, salary and other benefits, and career advancement opportunities~\cite{schuth2018recruiting, hyrynsalmi2019underrepresentation}. Still, as women are still underrepresented in the field, even the best diversity recruitment practices~\cite{dagan2023building} do not help if there is no population from where to recruit. Thus, universities and other higher education providers are also challenged to attract students of diverse backgrounds and genders to their programs. 

Some evidence shows that female students who do not consider a software career for themselves still value IT highly~\cite{leiviska2010attitudes}. Furthermore, the female students attending the ICT programmes seem to perceive it as a prospective profession, even if they may lack self-esteem compared to their male counterparts~\cite{rankovic2019female}. Diminishing gendered stereotypes in the ICT careers and underlining the required "soft" skills seem to be the first steps towards higher participation of women in the tech industry. Still, the evidence also suggests that these actions alone may not be enough~\cite{tam2020gender}. Even if the prospective female students see the opportunities in the ICT field, it may require more to get them actually apply to the ICT programmes. At the same time, the faculty and university teachers tend to see their programmes enough aware of diversity practices~\cite{hyrynsalmi2023how}. Still, the number of female students is increasing but too slowly.

\subsection{Where to find the female software engineers} 

The problem described here is long known, and successful projects have generated insights and ideas for more diverse curricula~\cite{travers2023shooting}. There have been hackathons~\cite{paganini2023opportunities}, actions towards women career-changers~\cite{hyrynsalmi2019underrepresentation, buhnova2019women} and other, often interdisciplinary, initiatives~\cite{happe2023rockstartit,garcia2019engaging} involving more female students in STEM and software engineering studies. These activities have a significant place in the field of raising awareness of technology professions. However, university student recruitment is an ongoing, long-term process different from specialising projects. It is often designed and implemented by specialised marketing professionals instead of STEM faculty whose work time is effectively reserved for research and education. Resourced used for marketing actions, especially in publicly funded universities (like in Finland, where this study takes place), are often limited and shared with all the other disciplines. These resource limitations often make hands-on activities and personal connections in the field very challenging.

University applicant attraction efforts can include multiple different marketing strategies~\cite{camilleri2020higher} and branding activities~\cite{beneke2011marketing,rauschnabel2016brand}. Physical recruitment, i.e. in-campus events and study fairs, is often reported as efficient applicant marketing together with the university websites and brochures~\cite{constantinides2011potential} and student-written web content~\cite{sandlin2014building} targeted for the applicants. In addition, universities and other higher education providers are attracted by branding a social media presence~\cite{constantinides2011potential} for marketing and recruitment efforts. These can include hosting university-specific pages or feeds~\cite{assimakopoulos2017effective} but also ads in video streaming services such as Youtube~\cite{constantinides2012higher} and even utilising services of social media influencers~\cite{schneckenleitner2020use,rolfs2021opportunities}. Social media campaigns often target younger generations through social media channels such as TikTok, Snapchat, and Instagram~\cite{shields2019social}. More "adult" channels such as LinkedIn may be dismissed~\cite{shields2019social} even if, as we later show in this paper, several students apply in their first or second university programme only later in life. The same risk lies with the student recruitment events as they are usually targeted at the local high school students.



\section{Research approach}\label{sec:research}
\subsection{Research process}
This study is carried out as a survey of university students majoring in software engineering at Finnish universities as this exploratory study aims to generate new insights into the topic. The chosen research strategy follows a mixed-method approach. 
%
The research process used in this study consists of two major phases: preparation and execution. The first major phase consists of literature and background study, identifying the relevant institutions in Finland, and contacting their education programme managers. 

First, after the literature study, we listed all universities in Finland that have a major in software engineering. In this phase of study, polytechnics were omitted. Out of the 13 universities in Finland, nine universities are offering degree studies in software engineering (Aalto University, LUT University, Tampere University, University of Eastern Finland, University of Helsinki, University of Jyväskylä, University of Oulu, University of Turku, and Åbo Akademi University). We contacted software engineering professors in each university and inquired about information regarding the persons responsible for marketing or leading software engineering programs in their institution. 

During August-September 2023, we held separate meetings with each university's responsible marketing programme managers. During the discussion, we confirmed their institutions' software engineering education programmes and invited them to forward our questionnaire to their students. All universities agreed to take part in the study.

The second major phase consisted of designing, conducting and analysing the survey of the software engineering major students. Each university's corresponding person sent the survey directly to the software engineering majors in their university in August-September 2023. Some universities also sent a remainder later, usually in mid-October. Finally, the analysis and reporting were carried out during November-December 2023.

\subsection{Survey design}

The survey instrument was designed in August 2023. The survey was designed to be exploratory in nature as it aims to generate insights into the topic at hand. The questions relevant to this study's goals were characterised and improved in meetings with the authors. The used questions include free answer questions, ranking questions, and question asking to select the most suitable option. The survey was implemented using the Webropol online questionnaire survey tool.

The selected 22 questions were grouped into three main thematical groups: \textit{(i)} experiences and reasons to select software engineering, \textit{(ii)} experiences on marketing activities, and \textit{(iii)} background information. The survey was designed to be anonymous; however, the basic background information (age, gender, etc.) was asked.

The survey was designed and implemented in Finnish as this research focused on the effects of marketing activities in Finland. While most of the studied universities offer either Bachelor's or Master's programmes taught in English, the marketing strategy and activities for those programmes often differ from the marketing to Finnish-speaking citizens, often applying for their first study position in a higher-education institution.   

The designed survey was tested with two students and reviewed by one software engineering programme leader. Based on their feedback, some questions were clarified and reworded to make them unambiguous. 

Soon after the study was published, it was noted that the university's website was a frequently appearing option- among the free answers for the questions where the respondents sought further information regarding the study programmes. This option was then added among the alternatives in the respective question. 

\subsection{Participant recruitment}

The survey is designed for university students majoring in software engineering at Finnish universities. Especially the focus is on understanding the impact of marketing activities focused on Finnish-speaking candidates and their influences on different genders. The study is designed as an open public subscription survey; that is, the objective is to gather as many respondents belonging to the target population as possible. 

The size of the target population is hard to estimate as the official role of the software engineering education programmes varies a lot between different institutions. However, according to statistics from The Ministry of Education and Culture and the Finnish National Agency for Education, there are 7,734 students in Finnish universities in different programmes in the field of Information Technology. This number, however, includes also, for instance, computer engineering, telecommunication, and data science majors. As a rough estimate, around 1,000--2,000 students major in software engineering. 

The participant recruitment was done via two main channels: First and foremost, each university's software engineering program leader, or corresponding, was asked to send the invitation to the students of their programmes. All universities agreed to do this. Second, the survey questionnaire was distributed through social media channels such as through various student organisations' X (Twitter), LinkedIn, and Instagram accounts. 

\subsection{Analysis}


The survey consisted of both qualitative and quantitative questions, and therefore, a mixed-method approach for the analysis was selected. Mixed methods are suitable for this study as the research objectives, on the one hand, involve both qualitative (e.g., RQ1) and quantitative (e.g., RQ2) aspects; in addition, the research approach in this study is exploratory, thus benefiting from a mixed-method approach.   

The quantitative answers and the corresponding open answers were considered to provide qualitative insights into the numbers. Quantitative answers were divided into female and male groups, and the statistical independence between these groups was tested with suitable statistical tests. The tests used were the Wilcoxon rank sum test, which is a nonparametric test for two populations when samples are independent, and the Kruskal-Wallis test, which is a nonparametric version of classical one-way ANOVA and an extension of the Wilcoxon rank sum test to more than two groups. 

For the qualitative analysis, we used thematic analysis as guided by~~\citeauthor{braun2006using}~~\cite{braun2006using}. We included the 13 open-ended questions used in the survey for the qualitative analysis. The analysis was done in the six steps: \textit{(i)} familiarising with the data, \textit{(ii)} initial codes, \textit{(iii)} theme identification, \textit{(iv)} reviewing themes, \textit{(v)} naming themes, and \textit{(vi)} reporting. Finally, the qualitative and quantitative analysis results are combined into a simplified model.

\section{Results}\label{sec:results}


\subsection{Demographic distribution}

The demographic information was collected in the following categories. Gender was requested as a categorical variable female, male, other, and do not wish to disclose. Similarly, the age group was a categorical variable in decades, i.e. 18 -- 20 years old as a fresh high school graduate, 21 -- 30 years old as a little older student, and so on. In addition, previous work experience, previously completed degrees, study degree level (both BSc and MSc or MSc programme options) and the name of the home university were requested. 

The questionnaire received 78 answers, from which 38 participants identified as a woman (50.7\%), 35 as a male (46.7\%), one other, and one did not prefer to disclose. Due to only two participants reporting something other than female or male, this study focuses on these two gender categories only and later discusses only participants identifying as female or male (N=76). The age distribution is given in Table \ref{tab:ages}, showing that, on average, the male participants were younger than the female participants. Around half of the participants had no previous work experience before beginning their university degree. 32.4\% of women had previous ICT-related work experience and 21.6\% some other type of work experience. Similarly, 37.2\% of males had ICT-related work experience, and 11.4\% had some other field experience. 
	
63.9\% of female participants and 68.6\% of male participants had a study right for both BSc and MSc degrees. 36.1\% of female participants and 31.4\% of male participants only studied in an MSc programme, usually indicating that this is their second degree (either completed BSc in another programme or in a university of applied sciences. The questionnaire was sent to students of nine national universities where software engineering (or a similar field with a different name) is taught. The respondents came from seven different universities.

\begin{table}[]
    \centering
    \begin{tabular}{l | l l}
    & Women & Male \\ \hline
18-20 & 5.3\% & 5.7\% \\
21-30 & 23.7\% & \textbf{54.3\%} \\
31-40 & \textbf{52.6\%} & 28.6\% \\
41-50 & 15.8\% & 8.6\% \\
51-60 & 2.6\% & 2.8\%
    \end{tabular}
    \caption{Age distribution of the questionnaire participants by gender. The major age groups are highlighted. All answers are between 18 and 60 years old.}
    \label{tab:ages}
\end{table}

\subsection{When do they apply for the software engineering studies?}



\textbf{"Do you remember when you decided to apply to study software engineering?"} This question aims to survey when students first thought of studying the ICT field, or software engineering more specifically, compared to their respective school history. The options were given as primary school (age 7 to 12 in the Finnish school system), secondary school (age 13 to 15), high school (or corresponding vocational school, typically age 15 to 18), university of applied sciences (after age 18; often an option for the university studies in the first place), or later in the adulthood, after finishing the first formal education. The option "other" was given but received only an answer "during the gap year". 

The results of this question are shown in Figure \ref{fig:when}. The answers are divided by gender. Our analysis with the Wilcoxon rank sum test shows that there is a statistical difference between female and male groups ($p = 4.3248e-23$); female participants decided to achieve software engineering degrees later in their lives, whereas male participants already figured their future field in the high school age. Our study attracted many participants whose decision to study software engineering was made later in life, which is also visible in the age distribution (Table \ref{tab:ages}). Especially many female participants belong to the age group 31--40 years old; at the same time, 68.4\% of our female participants decided to achieve a degree in the field only after their first formal education. 


In open responses, students highlighted different reasons for applying to software engineering education and their motivation for it. For example, some female respondents reported a longstanding interest in technology. Still, for various reasons, they were guided towards different fields in their studies, and only later did they decide to pursue software engineering. This deviation could have been due to a lack of study counselling or perhaps the influence of a family member:

\begin{figure}
    \centering
    \includegraphics[width=\linewidth]{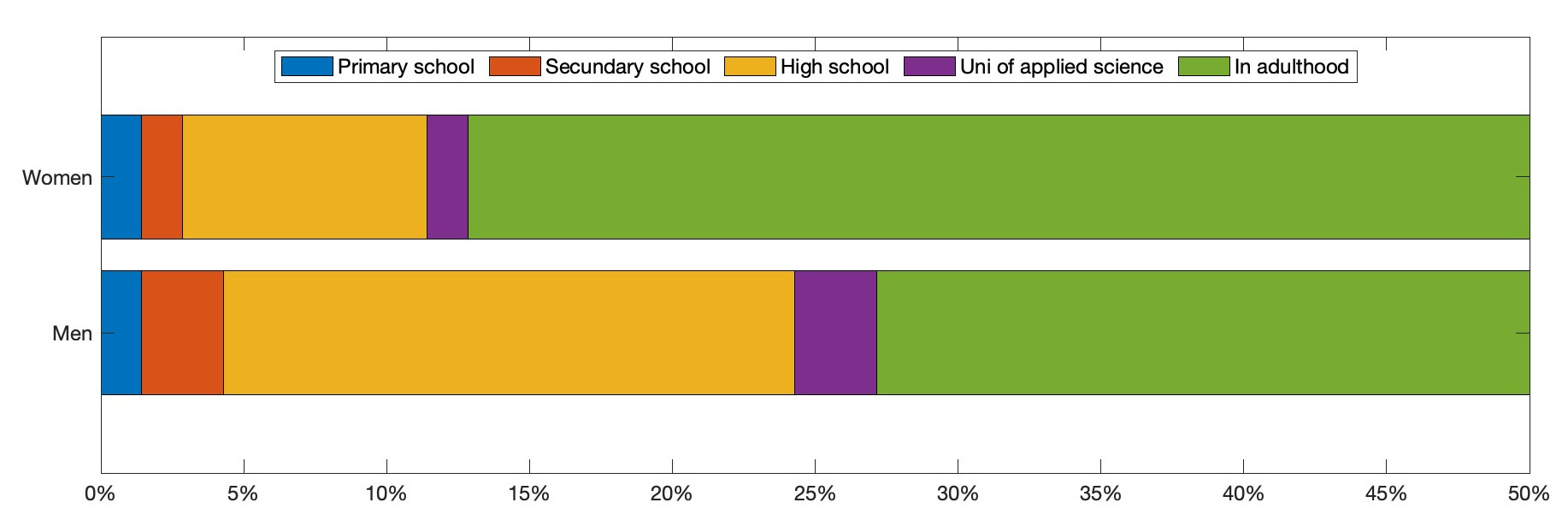}
    \caption{Answer distribution to the question "Do you remember when you decided to apply to study computer science?" (p = 4.3248e-23)}
    \label{fig:when}
\end{figure}

\textit{"For me, the main reason to study computer science was a very long-standing desire to pursue a technical degree. I considered applying for IT studies in high school, but my guidance counsellor and family encouraged me apply for business studies."} Woman, 31-40 years old

Some respondents illustrated a journey of rediscovery and overcoming initial apprehensions about software engineering. Although some students had previously considered software engineering to be an interesting field, they initially believed the studies to be too challenging and assumed coding skills as a prerequisite. However, their exposure to IT work in a different professional setting reignited their interest and led them to reconsider software engineering as a viable and desirable educational path:

\textit{"My interest was sparked while working in a different field when I had the opportunity to see up close the work my colleagues in IT were doing. Software Engineering had already crossed my mind, but I thought the studies would be too difficult then and that practical coding skills were required beforehand."} Man, 21-30 years old






\subsection{What influences on the decision to apply?}

\begin{figure}
    \centering
    \includegraphics[width=0.9\linewidth]{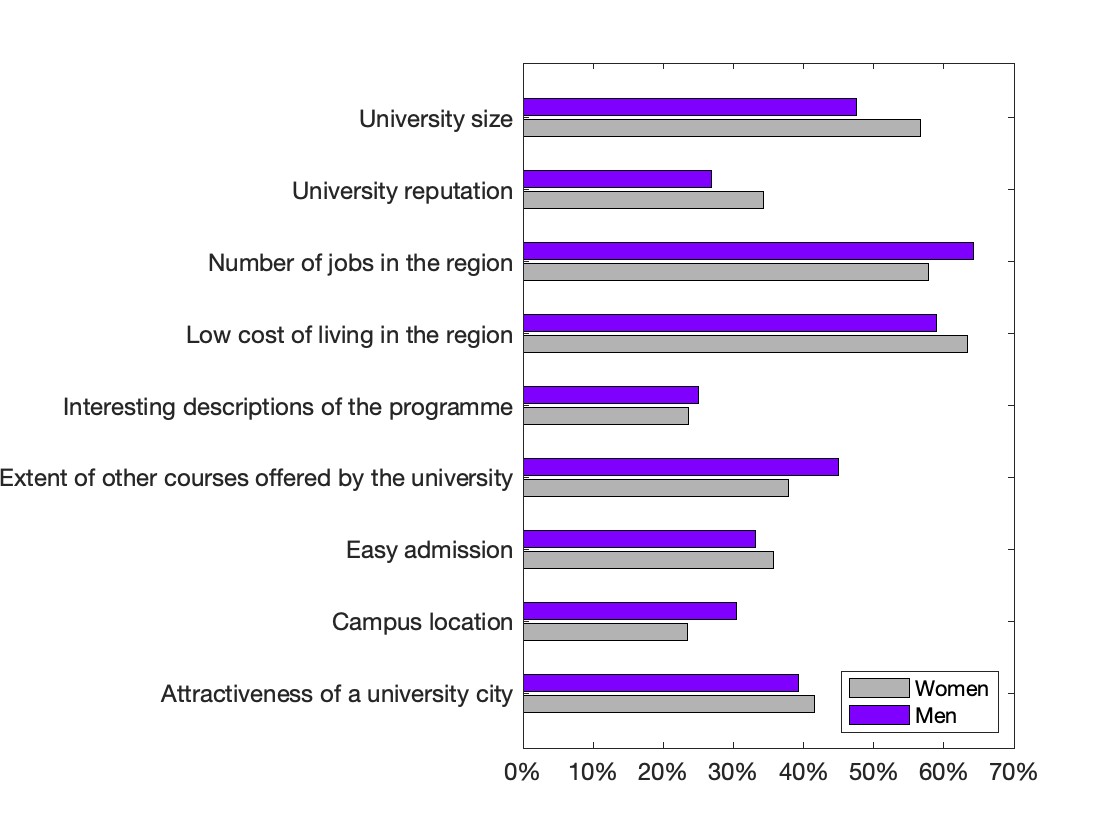}
    \caption{Answer distribution to the question "What influenced you to apply to your university?" (p = 0.0073)}
    \label{fig:vaikuttavuus}
\end{figure}

\textbf{Where did you find out about the programme you were accepted into? What influenced you to apply to your university?} This question aims to understand the influence of different pre-defined features on the decision-making of where to apply (programme or university). The question was answered by ranking suitable options by importance. Not all the options had to be picked, and an open answer was provided (discussed later in this paper). 

Figure \ref{fig:vaikuttavuus} summarises the results as average grades for each option, divided into gender-specific categories. The statistical analysis with the Kruskal-Wallis test returns a statistical difference between the answers of female and male participant groups ($p = 0.0073$). However, the differences are subtle: both gender categories preferred similar topics, such as the cheap cost of living in the university area, the number of job opportunities in the region, and the university's size in their answers, only with different weights. 

In the open responses, students revealed more information about their decision-making process. The role of social media was pointed out in many answers. Respondents noted seeing posts on platforms like Instagram and Facebook, and LinkedIn profiles of current students, which highlighted the professional opportunities available post-graduation:

\textit{"I saw posts on Instagram and Facebook. Additionally, I saw people on LinkedIn.com who study in the same program and the kinds of jobs they have."} Woman, 31-40 year old

Key factors influencing the decision to enrol on some programs included the program's inclusive eligibility criteria, allowing applications from individuals with diverse academic backgrounds, not limited to technical fields. Additionally, the possibility for remote studies and the flexibility of a schedule-independent learning format were cited as significant advantages:

\textit{"The primary reasons for choosing are: 1. Suitable threshold criteria for master's selection -> can apply with a lower academic background, not from a technical field. 2. Studies are handled entirely remotely. 3. Studies are almost entirely independent of schedule."} Woman, 41-50 years old






\subsection{What was the influence of university marketing on application decisions?}

\begin{figure}
    \centering
    \includegraphics[width=0.9\linewidth]{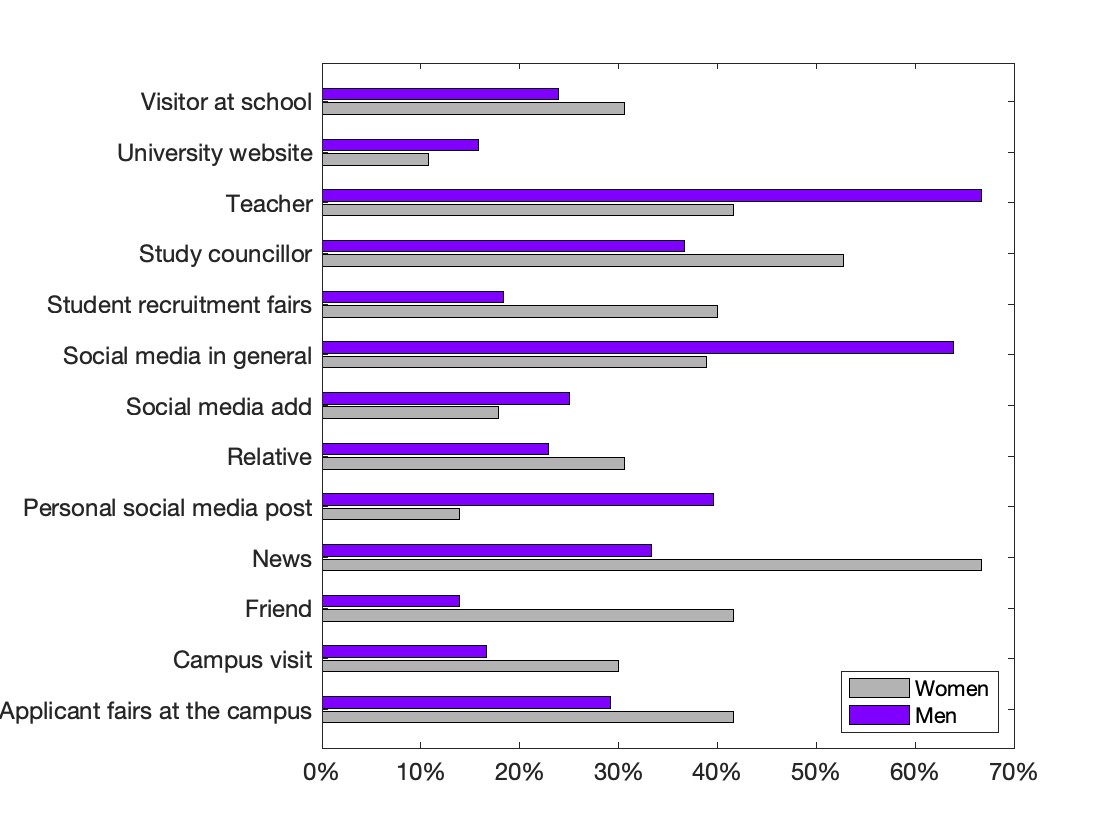}
    \caption{Answer distribution to the question "Where did you find out about the programme you got into?" (p = 0.0137)}
    \label{fig:markkinointi}
\end{figure}

\textbf{Where did you find out about the programme you got into?} This question aims to find out resources where the current students gathered information about their upcoming studies. The pre-defined features on the information sources were given as different types of visits (either into the university or different types of student fairs, or a visitor from the university coming into the school environment), social media (either an ad or more personal post), and other more close-tied communication forms such as hearing/discussing with a teacher, relative, or friend. Also, the University's presence in general news and the University's websites were included. The question was answered by ranking suitable options in order of importance. Not all the options had to be picked, and an open answer was provided (discussed later in this paper).

\begin{figure*}[t]
\centering
    \includegraphics[width=0.75\linewidth]{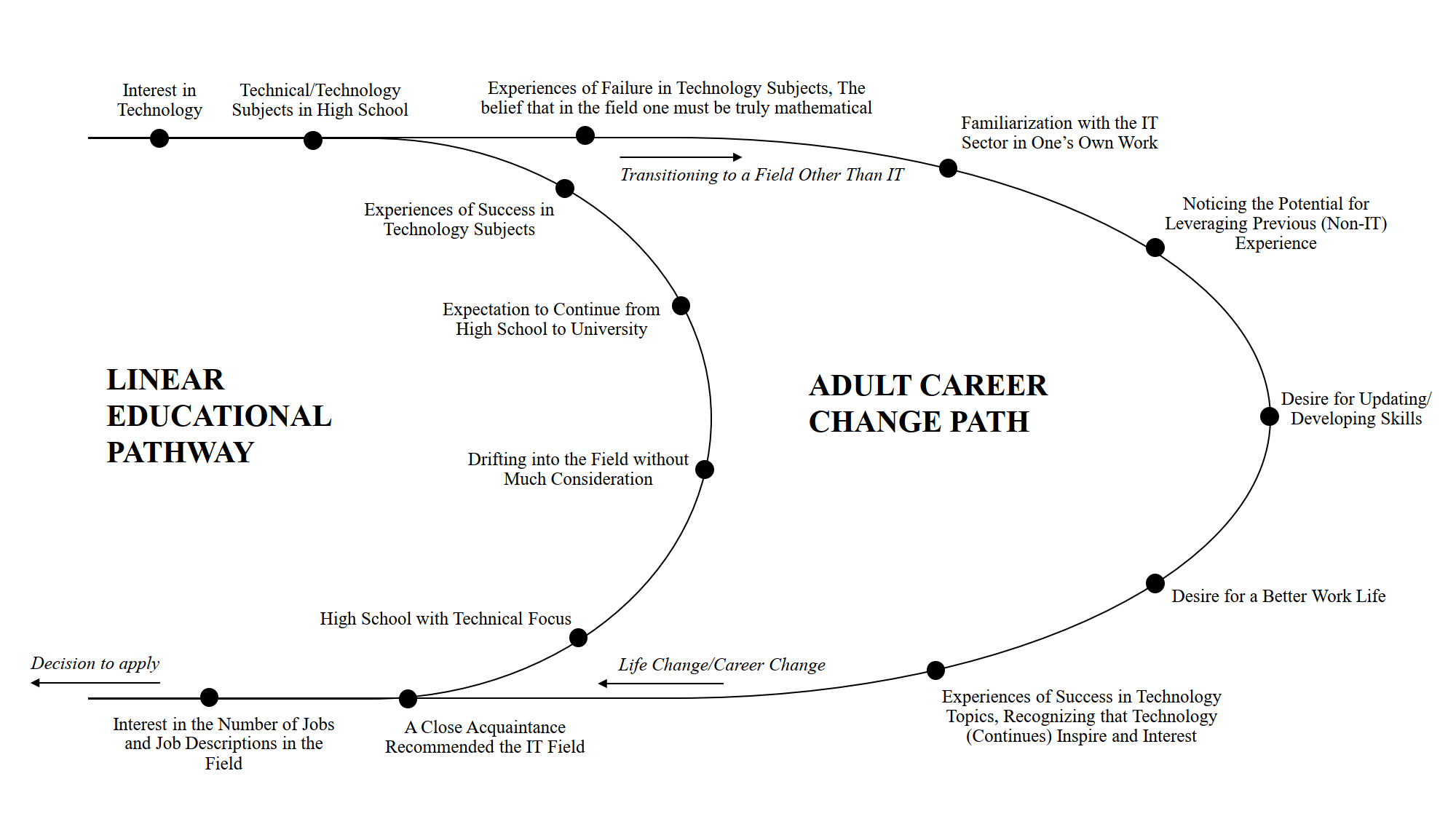}
    \caption{Diverse Paths to IT: Linear Educational Pathway and Adult Career Change Path identified in thematic analysis}
    \label{polku}
\end{figure*}

Figure \ref{fig:markkinointi} summarises the answers to this question, divided by gender. The Kruskal-Wallis test returns a statistical difference between the answers of female and male participant groups ($p = 0.0137$). 


Beyond effective social media marketing, the presence of easily accessible and experienced faculty members or role models was identified as crucial and influential. Typically, a single advertisement alone was not a decisive factor; instead, it was the combination of multi-channel and diverse voices in communication that contributed to making an informed decision:

\textit{"I first came across the advertisement for the program I'm in now in a Facebook group, where a faculty member of my university encouraged people to apply (the group was, if I recall correctly, a coder community group). This led me to explore the university's website and the specific program's page, and eventually, I contacted the faculty member directly before applying to the program."} Woman, 31-40 years old.

The influence of close relationships in decision-making was particularly evident among participants, both in close-ended questions and voluntary open-ended questions, when asked about finding information about the programme:

\textit{"I generally don't use social media and don't visit forums either. I try to block advertising by all means. A friend of mine previously studied at the same university, and I was interested in seeing what it's really like to study there. However, I gathered most of my information from the university's website."} Man, 31-40 years old

\subsection{The Impact of Education and Work Experience to the Path Towards IT}

In the thematic analysis of the answers to the questions "\textbf{How has your previous work experience influenced your chosen field and specialisation?}" and "\textbf{How has your educational journey shaped your career decisions so far?}", we revealed a multifaceted narrative. The results form a life-course~\cite{castaa2011understanding} inspired theory (Figure \ref{polku}) of the "Linear Educational Pathway" and "Adult Career Change Path" forming diverse paths towards software engineering education. 

Linear Educational Pathway presents a more 'traditional' path, where individuals move directly from the high school to the university. In thematic analysis, these respondents were divided into two groups: Those who had a passion for technology and were studying maybe in some high school with a technical focus. Then there are those who are more uncertain about where their passion lies but decide to apply for the study position at the university, as was expected of them.

\textit{"High school studies mainly influenced me in such a way that choosing this field did not feel too scary. I presumed that} [high school's] \textit{advanced mathematics} [syllabus] \textit{would provide a good foundation for succeeding in further studies."} Woman, 51-60 years old

In Linear Educational Pathway, the main drivers were positive experiences around technology. However, the thematic analysis revealed that experiences were also crucial for those who made career changes to software engineering in adult life --- the second round in this research. In analysis, experiences of failure in technology subjects or beliefs such as that the software engineering field requires deep mathematics occurred in the answers of adult career change path respondents. These respondents saw their past working experience as valuable for the IT sector, desired a better work-life (such as work with remote work possibilities or flexibility), or were looking for possibilities for personal growth and the possibility to update their skills. What was common was that they got positive experience with technology topics in their career, usually via their work duties around IT. Those positive experiences were strong enough to overcome previous bad experiences during high school and supported the career change to IT.

\textit{"I see that my previous study path had very little effect on my current choice because I come from a high school where studies in the field of technology were not really advertised at all for female people, and I ended up applying and getting a study place at a business school after high school. I think my choices have been influenced more by the strong role models I found in adulthood as well as my own non-technical career in the IT industry, which eventually made me realise that the IT industry and the studies in the field are really for everyone"} Woman, 31-40 years old

What was similar between these two paths was that they both shared an interest in technology, starting from a young age. In their decision to apply towards software engineering education, close acquaintances such as family members, teachers, or life partners played an important role. In these paths, respondents also reported that they were interested in pursuing software engineering, as there has been an ongoing discussion in Finland about the labour shortage problem in software engineering:

\textit{"(Study Path) Had no actual effect. The expert shortage in IT jobs has affected my decision much more."} Man, 41-50 years old

This theory of diverse paths to IT sheds light on how the interplay of professional experiences and educational background uniquely positions individuals in their career paths. It is essential to recognise that while our study identifies two primary pathways into IT studies – directly from high school and adult career change – these paths do not reflect the unique experiences of every individual. Both paths, despite their differences, reveal crucial points where paths divide from each other and in which points they share similar influence points. Related to the previous findings from the quantitative data, it may also be that these experiences of adult career change paths are more frequently the experiences of women. 

\section{Discussion}\label{sec:disc}

\subsection{Key findings}

We summarise our study's key findings in the following:
\begin{description}
    \item[First,] our study shows that, on average, women are applying later into software engineering studies, and the difference between male and female answers is statistically significant. 
    \item[Second,] the most important rationales behind the university selection for software engineering studies are the size of the university, cheap cost of living, and the number of job opportunities. In addition, the diverse job opportunities and the flexibility of the studies are seen as important.
    \item[Third,] the results show that the most affecting marketing actions to women are personal guidance; however, teachers and social media are the most important for men.
    \item[Fourth,] the results show two main, yet remarkably differentiating, paths to the field: the linear educational and adult career change paths. This concept helps to understand that there are several career paths in the field of software engineering, and they can be very different depending on gender and life cycle stages.
\end{description}

Our study supports the findings that women are interested in software engineering. Still, they may sometimes have to do a second round --- first, go to one career and then make a career change to software engineering. The findings of our thematic analysis also support the previous findings~\cite{hill2010so} that social and environmental factors contribute to the underrepresentation of women in engineering --- not their lack of interest in mathematical skills. The stereotype image that the software engineering field requires deep mathematical skills is still strong. 

The emphasis on the desire for a better work-life, flexibility, and remote learning options when making the decision points to a potential shift in priorities. This can be possibly influenced by post-COVID-19 trends around remote and hybrid working, which have been favourable to women and to increasing the proportion of women and reducing the workload in the home, and sometimes quite opposite~\cite{carli2020women}. This raises the question of whether there are barriers on the path to technical fields, especially for women, that could be mitigated through more flexible and accessible educational models. This could be one future study angle. 

Our study also reveals that personal guidance is crucial in attracting women to software engineering, while teachers and social media influence more men. This difference suggests a need for tailored outreach strategies. For women who may no longer have access to traditional educational guidance, such as high school counselling, in their adult career change path, alternative methods of personal outreach, like active engagement through various social media channels and events, could be effective. As a result that men favour social media, there can be a marketing bias: There have been previous findings that IT jobs were primarily targeted at men. In that way, more men clicked the advertisements and applied for the positions.~\cite{lambrecht2019algorithmic} This can also be an issue in our case. Furthermore, in our research, we found out that the presence of faculty members in social media spaces, both for program promotion and for providing accessible information about studies, is an effective channel and way to encourage women to apply. This indicates that it matters how social media is used to attract more women to software engineering --- but it also raises a question about resources, as this kind of marketing would require someone from the faculty.

Our study presents findings that add a new dimension to the ongoing discussion of attracting more women to software engineering education. Our 'Diverse Paths to IT' theory emphasises that universities' programme marketing strategies in the field of software engineering should identify and support both paths. Especially for women, the "adult career change path" is significant, as women in our research seek an IT career later in life, utilising their previous experiences and skills.




\subsection{Limitations and future studies}

Naturally, there are limitations worth discussing. Age distribution is skewed towards older students. As usual, those with strong opinions or the topic is close to their heart are more willing to answer the survey. This is also visible in our study, as we got half of the results from women, even if they represent only circa 25\% of Finnish university students. We will consider this in future studies and repeat the questionnaire with a larger population. This moment study focus only to the Finnish students, but can be repeated in the global perspective later. The questionnaire was only shared with the student population that was indeed accepted into the university studies; it would be interesting to study those who applied to the software engineering programs but did not get accepted.


\section{Conclusion}\label{sec:conc}
This study addressed two research questions: why did Finnish software engineering students select their current major, and how did they decide to apply to study software engineering? In addition, we studied whether there are differences between genders in these two questions. The survey results (n=78) showed that women often decided later in their life to apply to study in software engineering than men. In addition, there are differences in the main influencing factor driving the software engineering field; however, this might be partially explained by the age differences between the male and female respondents. We summarised the key findings into a model depicting two different paths to software engineering studies: the linear path and the career change path. The latter of these explains the name of this study: for a remarkable number of students in Finnish universities, software engineering studies are the second round after completing the first degree in some other field.  

\begin{acks}
This research was supported by PHP Säätiö (PHP Holding), Grant 20230012 
\end{acks}


\begin{thebibliography}{48}


\ifx \showCODEN    \undefined \def \showCODEN     #1{\unskip}     \fi
\ifx \showDOI      \undefined \def \showDOI       #1{#1}\fi
\ifx \showISBNx    \undefined \def \showISBNx     #1{\unskip}     \fi
\ifx \showISBNxiii \undefined \def \showISBNxiii  #1{\unskip}     \fi
\ifx \showISSN     \undefined \def \showISSN      #1{\unskip}     \fi
\ifx \showLCCN     \undefined \def \showLCCN      #1{\unskip}     \fi
\ifx \shownote     \undefined \def \shownote      #1{#1}          \fi
\ifx \showarticletitle \undefined \def \showarticletitle #1{#1}   \fi
\ifx \showURL      \undefined \def \showURL       {\relax}        \fi
\providecommand\bibfield[2]{#2}
\providecommand\bibinfo[2]{#2}
\providecommand\natexlab[1]{#1}
\providecommand\showeprint[2][]{arXiv:#2}

\bibitem[Acilar and S{\ae}b{\o}(2022)]%
        {acilar2022exploring}
\bibfield{author}{\bibinfo{person}{Ali Acilar} {and} \bibinfo{person}{{\O}ystein S{\ae}b{\o}}.} \bibinfo{year}{2022}\natexlab{}.
\newblock \showarticletitle{Exploring the differences between students in IS and other disciplines in the perceptions of factors influencing study program choice: A survey study in Norway}.
\newblock  (\bibinfo{year}{2022}).
\newblock


\bibitem[Albusays et~al\mbox{.}(2021)]%
        {albusays2021diversity}
\bibfield{author}{\bibinfo{person}{Khaled Albusays}, \bibinfo{person}{Pernille Bjorn}, \bibinfo{person}{Laura Dabbish}, \bibinfo{person}{Denae Ford}, \bibinfo{person}{Emerson Murphy-Hill}, \bibinfo{person}{Alexander Serebrenik}, {and} \bibinfo{person}{Margaret-Anne Storey}.} \bibinfo{year}{2021}\natexlab{}.
\newblock \showarticletitle{The diversity crisis in software development}.
\newblock \bibinfo{journal}{\emph{IEEE Software}} \bibinfo{volume}{38}, \bibinfo{number}{2} (\bibinfo{year}{2021}), \bibinfo{pages}{19--25}.
\newblock


\bibitem[Assimakopoulos et~al\mbox{.}(2017)]%
        {assimakopoulos2017effective}
\bibfield{author}{\bibinfo{person}{Costas Assimakopoulos}, \bibinfo{person}{Ioannis Antoniadis}, \bibinfo{person}{Oliver~G Kayas}, {and} \bibinfo{person}{Dragana Dvizac}.} \bibinfo{year}{2017}\natexlab{}.
\newblock \showarticletitle{Effective social media marketing strategy: Facebook as an opportunity for universities}.
\newblock \bibinfo{journal}{\emph{Int. Journal of Retail \& Distribution Management}} \bibinfo{volume}{45}, \bibinfo{number}{5} (\bibinfo{year}{2017}).
\newblock


\bibitem[Beneke(2011)]%
        {beneke2011marketing}
\bibfield{author}{\bibinfo{person}{JH Beneke}.} \bibinfo{year}{2011}\natexlab{}.
\newblock \showarticletitle{Marketing the institution to prospective students-A review of brand (reputation) management in higher education}.
\newblock \bibinfo{journal}{\emph{Int. journal of Business and Management}} \bibinfo{volume}{6}, \bibinfo{number}{1} (\bibinfo{year}{2011}), \bibinfo{pages}{29}.
\newblock


\bibitem[Bottia et~al\mbox{.}(2015)]%
        {bottia2015growing}
\bibfield{author}{\bibinfo{person}{Martha~Cecilia Bottia}, \bibinfo{person}{Elizabeth Stearns}, \bibinfo{person}{Roslyn~Arlin Mickelson}, \bibinfo{person}{Stephanie Moller}, {and} \bibinfo{person}{Lauren Valentino}.} \bibinfo{year}{2015}\natexlab{}.
\newblock \showarticletitle{Growing the roots of STEM majors: Female math and science high school faculty and the participation of students in STEM}.
\newblock \bibinfo{journal}{\emph{Economics of Education Review}}  \bibinfo{volume}{45} (\bibinfo{year}{2015}), \bibinfo{pages}{14--27}.
\newblock


\bibitem[Braun and Clarke(2006)]%
        {braun2006using}
\bibfield{author}{\bibinfo{person}{Virginia Braun} {and} \bibinfo{person}{Victoria Clarke}.} \bibinfo{year}{2006}\natexlab{}.
\newblock \showarticletitle{Using thematic analysis in psychology}.
\newblock \bibinfo{journal}{\emph{Qualitative Research in Psychology}} \bibinfo{volume}{3}, \bibinfo{number}{2} (\bibinfo{year}{2006}), \bibinfo{pages}{77--101}.
\newblock


\bibitem[Buhnova and Prikrylova(2019)]%
        {buhnova2019women}
\bibfield{author}{\bibinfo{person}{Barbora Buhnova} {and} \bibinfo{person}{Dita Prikrylova}.} \bibinfo{year}{2019}\natexlab{}.
\newblock \showarticletitle{Women want to learn tech: Lessons from the Czechitas education project}. In \bibinfo{booktitle}{\emph{2019 IEEE/ACM 2nd International Workshop on Gender Equality in Software Engineering (GE)}}. IEEE, \bibinfo{pages}{25--28}.
\newblock


\bibitem[Camilleri(2020)]%
        {camilleri2020higher}
\bibfield{author}{\bibinfo{person}{Mark Camilleri}.} \bibinfo{year}{2020}\natexlab{}.
\newblock \showarticletitle{Higher education marketing communications in the digital era}.
\newblock In \bibinfo{booktitle}{\emph{Strategic marketing of Higher education in Africa}}. \bibinfo{publisher}{Routledge}, \bibinfo{pages}{77--95}.
\newblock


\bibitem[Canaan and Mouganie(2023)]%
        {canaan2023impact}
\bibfield{author}{\bibinfo{person}{Serena Canaan} {and} \bibinfo{person}{Pierre Mouganie}.} \bibinfo{year}{2023}\natexlab{}.
\newblock \showarticletitle{The impact of advisor gender on female students’ STEM enrollment and persistence}.
\newblock \bibinfo{journal}{\emph{Journal of Human Resources}} \bibinfo{volume}{58}, \bibinfo{number}{2} (\bibinfo{year}{2023}), \bibinfo{pages}{593--632}.
\newblock


\bibitem[Carli(2020)]%
        {carli2020women}
\bibfield{author}{\bibinfo{person}{Linda~L Carli}.} \bibinfo{year}{2020}\natexlab{}.
\newblock \showarticletitle{Women, gender equality and COVID-19}.
\newblock \bibinfo{journal}{\emph{Gender in Management: An Int.Journal}} \bibinfo{volume}{35}, \bibinfo{number}{7/8} (\bibinfo{year}{2020}), \bibinfo{pages}{647--655}.
\newblock


\bibitem[Castaño and Webster(2011)]%
        {castaa2011understanding}
\bibfield{author}{\bibinfo{person}{C.~Castaño} {and} \bibinfo{person}{J.~Webster}.} \bibinfo{year}{2011}\natexlab{}.
\newblock \showarticletitle{Understanding women’s presence in ICT: The life course perspective}.
\newblock \bibinfo{journal}{\emph{Int. J. Gender, Science \& Technology}} \bibinfo{volume}{3}, \bibinfo{number}{2} (\bibinfo{year}{2011}), \bibinfo{pages}{364--386}.
\newblock


\bibitem[Catolino et~al\mbox{.}(2019)]%
        {catolino2019gender}
\bibfield{author}{\bibinfo{person}{Gemma Catolino}, \bibinfo{person}{Fabio Palomba}, \bibinfo{person}{Damian~A Tamburri}, \bibinfo{person}{Alexander Serebrenik}, {and} \bibinfo{person}{Filomena Ferrucci}.} \bibinfo{year}{2019}\natexlab{}.
\newblock \showarticletitle{Gender diversity and women in software teams: How do they affect community smells?}. In \bibinfo{booktitle}{\emph{IEEE/ACM 41st Int. Conf. on Software Engineering: Software Engineering in Society (ICSE-SEIS)}}. IEEE, \bibinfo{pages}{11--20}.
\newblock


\bibitem[Constantinides and Stagno(2012)]%
        {constantinides2012higher}
\bibfield{author}{\bibinfo{person}{E.~Constantinides} {and} \bibinfo{person}{M.C~Zinck Stagno}.} \bibinfo{year}{2012}\natexlab{}.
\newblock \showarticletitle{Higher education marketing: A study on the impact of social media on study selection and university choice}.
\newblock \bibinfo{journal}{\emph{Int. journal of technology and educational marketing (IJTEM)}} \bibinfo{volume}{2}, \bibinfo{number}{1} (\bibinfo{year}{2012}), \bibinfo{pages}{41--58}.
\newblock


\bibitem[Constantinides and Zinck~Stagno(2011)]%
        {constantinides2011potential}
\bibfield{author}{\bibinfo{person}{Efthymios Constantinides} {and} \bibinfo{person}{Marc~C Zinck~Stagno}.} \bibinfo{year}{2011}\natexlab{}.
\newblock \showarticletitle{Potential of the social media as instruments of higher education marketing: A segmentation study}.
\newblock \bibinfo{journal}{\emph{Journal of marketing for higher education}} \bibinfo{volume}{21}, \bibinfo{number}{1} (\bibinfo{year}{2011}), \bibinfo{pages}{7--24}.
\newblock


\bibitem[Dagan et~al\mbox{.}(2023)]%
        {dagan2023building}
\bibfield{author}{\bibinfo{person}{Ella Dagan}, \bibinfo{person}{Anita Sarma}, \bibinfo{person}{Alison Chang}, \bibinfo{person}{Sarah D’Angelo}, \bibinfo{person}{Jill Dicker}, {and} \bibinfo{person}{Emerson Murphy-Hill}.} \bibinfo{year}{2023}\natexlab{}.
\newblock \showarticletitle{Building and Sustaining Ethnically, Racially, and Gender Diverse Software Engineering Teams: A Study at Google}. In \bibinfo{booktitle}{\emph{31st ACM Joint European Software Engineering Conference and Symposium on the Foundations of Software Engineering}} \emph{(\bibinfo{series}{ESEC/FSE 2023})}. \bibinfo{publisher}{Association for Computing Machinery}, \bibinfo{address}{New York, NY, USA}, \bibinfo{pages}{631–643}.
\newblock
\showISBNx{9798400703270}


\bibitem[Donmez(2021)]%
        {donmez2021impact}
\bibfield{author}{\bibinfo{person}{Ismail Donmez}.} \bibinfo{year}{2021}\natexlab{}.
\newblock \showarticletitle{Impact of Out-of-School STEM Activities on STEM Career Choices of Female Students.}
\newblock \bibinfo{journal}{\emph{Eurasian Journal of educational research}}  \bibinfo{volume}{91} (\bibinfo{year}{2021}).
\newblock


\bibitem[Ertl et~al\mbox{.}(2017)]%
        {ertl2017impact}
\bibfield{author}{\bibinfo{person}{Bernhard Ertl}, \bibinfo{person}{Silke Luttenberger}, {and} \bibinfo{person}{Manuela Paechter}.} \bibinfo{year}{2017}\natexlab{}.
\newblock \showarticletitle{The impact of gender stereotypes on the self-concept of female students in STEM subjects with an under-representation of females}.
\newblock \bibinfo{journal}{\emph{Frontiers in psychology}}  \bibinfo{volume}{8} (\bibinfo{year}{2017}), \bibinfo{pages}{703}.
\newblock


\bibitem[Fisher et~al\mbox{.}(2020)]%
        {fisher2020gender}
\bibfield{author}{\bibinfo{person}{Camilla~R Fisher}, \bibinfo{person}{Christopher~D Thompson}, {and} \bibinfo{person}{Rowan~H Brookes}.} \bibinfo{year}{2020}\natexlab{}.
\newblock \showarticletitle{Gender differences in the Australian undergraduate STEM student experience: a systematic review}.
\newblock \bibinfo{journal}{\emph{Higher Education Research \& Development}} \bibinfo{volume}{39}, \bibinfo{number}{6} (\bibinfo{year}{2020}), \bibinfo{pages}{1155--1168}.
\newblock


\bibitem[Garc{\'\i}a-Holgado et~al\mbox{.}(2019)]%
        {garcia2019engaging}
\bibfield{author}{\bibinfo{person}{Alicia Garc{\'\i}a-Holgado}, \bibinfo{person}{Amparo~Camacho D{\'\i}az}, {and} \bibinfo{person}{Francisco~J Garc{\'\i}a-Pe{\~n}alvo}.} \bibinfo{year}{2019}\natexlab{}.
\newblock \showarticletitle{Engaging women into STEM in Latin America: W-STEM project}. In \bibinfo{booktitle}{\emph{7th Int. Conf. on Technological Ecosystems for Enhancing Multiculturality}}. \bibinfo{pages}{232--239}.
\newblock


\bibitem[Gonz{\'a}lez-P{\'e}rez et~al\mbox{.}(2020)]%
        {gonzalez2020girls}
\bibfield{author}{\bibinfo{person}{Susana Gonz{\'a}lez-P{\'e}rez}, \bibinfo{person}{Ruth Mateos~de Cabo}, {and} \bibinfo{person}{Milagros S{\'a}inz}.} \bibinfo{year}{2020}\natexlab{}.
\newblock \showarticletitle{Girls in STEM: Is it a female role-model thing?}
\newblock \bibinfo{journal}{\emph{Frontiers in psychology}}  \bibinfo{volume}{11} (\bibinfo{year}{2020}), \bibinfo{pages}{2204}.
\newblock


\bibitem[Happe and Marquardt(2023)]%
        {happe2023rockstartit}
\bibfield{author}{\bibinfo{person}{Lucia Happe} {and} \bibinfo{person}{Kai Marquardt}.} \bibinfo{year}{2023}\natexlab{}.
\newblock \showarticletitle{RockStartIT: Authentic and Inclusive Interdisciplinary Software Engineering Courses}. In \bibinfo{booktitle}{\emph{2023 IEEE/ACM 4th Workshop on Gender Equity, Diversity, and Inclusion in Software Engineering (GEICSE)}}. IEEE.
\newblock


\bibitem[Hill et~al\mbox{.}(2010)]%
        {hill2010so}
\bibfield{author}{\bibinfo{person}{Catherine Hill}, \bibinfo{person}{Christianne Corbett}, {and} \bibinfo{person}{Andresse St~Rose}.} \bibinfo{year}{2010}\natexlab{}.
\newblock \bibinfo{booktitle}{\emph{Why so few? Women in science, technology, engineering, and mathematics.}}
\newblock \bibinfo{publisher}{ERIC}.
\newblock


\bibitem[Hyrynsalmi and Hyrynsalmi(2019)]%
        {hyrynsalmi2019software}
\bibfield{author}{\bibinfo{person}{Sonja Hyrynsalmi} {and} \bibinfo{person}{Sami Hyrynsalmi}.} \bibinfo{year}{2019}\natexlab{}.
\newblock \showarticletitle{Software engineering studies attractiveness for the highly educated women planning to change career in finland}. In \bibinfo{booktitle}{\emph{2019 IEEE/ACM 41st International Conference on Software Engineering: Companion Proceedings (ICSE-Companion)}}. IEEE, \bibinfo{pages}{304--305}.
\newblock


\bibitem[Hyrynsalmi(2019)]%
        {hyrynsalmi2019underrepresentation}
\bibfield{author}{\bibinfo{person}{Sonja~M Hyrynsalmi}.} \bibinfo{year}{2019}\natexlab{}.
\newblock \showarticletitle{The underrepresentation of women in the software industry: thoughts from career-changing women}. In \bibinfo{booktitle}{\emph{2019 IEEE/ACM 2nd International Workshop on Gender Equality in Software Engineering (GE)}}. IEEE.
\newblock


\bibitem[Hyrynsalmi(2023)]%
        {hyrynsalmi2023how}
\bibfield{author}{\bibinfo{person}{Sonja~M. Hyrynsalmi}.} \bibinfo{year}{2023}\natexlab{}.
\newblock \showarticletitle{How Diversity and Inclusion Are Approached in Software Engineering University-level Teaching}. In \bibinfo{booktitle}{\emph{IEEE/ACM 4th Workshop on Gender Equity, Diversity, and Inclusion in Software Engineering (GEICSE)}}. \bibinfo{pages}{17--24}.
\newblock


\bibitem[Kelly et~al\mbox{.}(2019)]%
        {kelly2019stem}
\bibfield{author}{\bibinfo{person}{Regina Kelly}, \bibinfo{person}{Oliver McGarr}, \bibinfo{person}{Louise Lehane}, {and} \bibinfo{person}{Sibel Erduran}.} \bibinfo{year}{2019}\natexlab{}.
\newblock \showarticletitle{STEM and gender at university: focusing on Irish undergraduate female students’ perceptions}.
\newblock \bibinfo{journal}{\emph{J. of Applied Research in Higher Education}} \bibinfo{volume}{11}, \bibinfo{number}{4} (\bibinfo{year}{2019}), \bibinfo{pages}{770--787}.
\newblock


\bibitem[Kohl and Prikladnicki(2022)]%
        {kohl2022benefits}
\bibfield{author}{\bibinfo{person}{Karina Kohl} {and} \bibinfo{person}{Rafael Prikladnicki}.} \bibinfo{year}{2022}\natexlab{}.
\newblock \showarticletitle{Benefits and Difficulties of Gender Diversity on Software Development Teams: A Qualitative Study}. In \bibinfo{booktitle}{\emph{XXXVI Brazilian Symposium on Software Engineering}}. \bibinfo{pages}{21--30}.
\newblock


\bibitem[Lambrecht and Tucker(2019)]%
        {lambrecht2019algorithmic}
\bibfield{author}{\bibinfo{person}{Anja Lambrecht} {and} \bibinfo{person}{Catherine Tucker}.} \bibinfo{year}{2019}\natexlab{}.
\newblock \showarticletitle{Algorithmic bias? An empirical study of apparent gender-based discrimination in the display of STEM career ads}.
\newblock \bibinfo{journal}{\emph{Management science}} \bibinfo{volume}{65}, \bibinfo{number}{7} (\bibinfo{year}{2019}), \bibinfo{pages}{2966--2981}.
\newblock


\bibitem[Leivisk{\"a} and Siponen(2010)]%
        {leiviska2010attitudes}
\bibfield{author}{\bibinfo{person}{Katja Leivisk{\"a}} {and} \bibinfo{person}{Mikko Siponen}.} \bibinfo{year}{2010}\natexlab{}.
\newblock \showarticletitle{Attitudes of sixth form female students toward the IT field}.
\newblock \bibinfo{journal}{\emph{ACM SIGCAS Computers and Society}} \bibinfo{volume}{40}, \bibinfo{number}{1} (\bibinfo{year}{2010}), \bibinfo{pages}{34--49}.
\newblock


\bibitem[Margolis and Fisher(2002)]%
        {margolis2002unlocking}
\bibfield{author}{\bibinfo{person}{Jane Margolis} {and} \bibinfo{person}{Allan Fisher}.} \bibinfo{year}{2002}\natexlab{}.
\newblock \bibinfo{booktitle}{\emph{Unlocking the clubhouse: Women in computing}}.
\newblock \bibinfo{publisher}{MIT press}.
\newblock


\bibitem[Milgram(2011)]%
        {milgram2011recruit}
\bibfield{author}{\bibinfo{person}{Donna Milgram}.} \bibinfo{year}{2011}\natexlab{}.
\newblock \showarticletitle{How to recruit women and girls to the science, technology, engineering, and math (STEM) classroom}.
\newblock \bibinfo{journal}{\emph{Technology and engineering teacher}} \bibinfo{volume}{71}, \bibinfo{number}{3} (\bibinfo{year}{2011}).
\newblock


\bibitem[Paganini et~al\mbox{.}(2023)]%
        {paganini2023opportunities}
\bibfield{author}{\bibinfo{person}{Lav{\'\i}nia Paganini}, \bibinfo{person}{Kiev Gama}, \bibinfo{person}{Alexander Nolte}, {and} \bibinfo{person}{Alexander Serebrenik}.} \bibinfo{year}{2023}\natexlab{}.
\newblock \showarticletitle{Opportunities and constraints of women-focused online hackathons}. In \bibinfo{booktitle}{\emph{IEEE/ACM 4th Workshop on Gender Equity, Diversity, and Inclusion in Software Engineering (GEICSE)}}. IEEE, \bibinfo{pages}{33--40}.
\newblock


\bibitem[Rajala et~al\mbox{.}(2022)]%
        {rajala2022perceiving}
\bibfield{author}{\bibinfo{person}{J.M.~Rajala}, \bibinfo{person}{N.~Iivari}, \bibinfo{person}{M.~Kinnula}, \bibinfo{person}{D.~Rajanen}, {and} \bibinfo{person}{T.~Molin-Juustila}.} \bibinfo{year}{2022}\natexlab{}.
\newblock \showarticletitle{Perceiving ICT: Factors Influencing the Selection of Information Systems as a Major}.
\newblock


\bibitem[Rankovi{\'c} et~al\mbox{.}(2019)]%
        {rankovic2019female}
\bibfield{author}{\bibinfo{person}{Nevena Rankovi{\'c}}, \bibinfo{person}{Elinda~Kajo Mece}, \bibinfo{person}{Mirjana Ivanovi{\'c}}, \bibinfo{person}{Asya Stoyanova-Doycheva}, \bibinfo{person}{Milo{\v{s}} Savi{\'c}}, {and} \bibinfo{person}{Dragica Rankovi{\'c}}.} \bibinfo{year}{2019}\natexlab{}.
\newblock \showarticletitle{Female Students' Attitude Towards Studying Informatics and Expectations for Future Career-Balkan Case}. In \bibinfo{booktitle}{\emph{9th Balkan Conference on Informatics}}. \bibinfo{pages}{1--7}.
\newblock


\bibitem[Rauschnabel et~al\mbox{.}(2016)]%
        {rauschnabel2016brand}
\bibfield{author}{\bibinfo{person}{Philipp~A Rauschnabel}, \bibinfo{person}{Nina Krey}, \bibinfo{person}{Barry~J Babin}, {and} \bibinfo{person}{Bjoern~S Ivens}.} \bibinfo{year}{2016}\natexlab{}.
\newblock \showarticletitle{Brand management in higher education: the university brand personality scale}.
\newblock \bibinfo{journal}{\emph{Journal of Business Research}} \bibinfo{volume}{69}, \bibinfo{number}{8} (\bibinfo{year}{2016}), \bibinfo{pages}{3077--3086}.
\newblock


\bibitem[Rolfs and Ahlquist(2021)]%
        {rolfs2021opportunities}
\bibfield{author}{\bibinfo{person}{Megan Rolfs} {and} \bibinfo{person}{Josie Ahlquist}.} \bibinfo{year}{2021}\natexlab{}.
\newblock \showarticletitle{Opportunities for student influencer marketing strategies}.
\newblock \bibinfo{journal}{\emph{Journal of Education Advancement \& Marketing}} \bibinfo{volume}{5}, \bibinfo{number}{4} (\bibinfo{year}{2021}).
\newblock


\bibitem[Russo and Stol(2020)]%
        {russo2020gender}
\bibfield{author}{\bibinfo{person}{Daniel Russo} {and} \bibinfo{person}{Klaas-Jan Stol}.} \bibinfo{year}{2020}\natexlab{}.
\newblock \showarticletitle{Gender differences in personality traits of software engineers}.
\newblock \bibinfo{journal}{\emph{IEEE Trans. on Software Engineering}} \bibinfo{volume}{48}, \bibinfo{number}{3} (\bibinfo{year}{2020}), \bibinfo{pages}{819--834}.
\newblock


\bibitem[Sandlin and Pe{\~n}a(2014)]%
        {sandlin2014building}
\bibfield{author}{\bibinfo{person}{J.K.~Sandlin} {and} \bibinfo{person}{E.~Vallejo Pe{\~n}a}.} \bibinfo{year}{2014}\natexlab{}.
\newblock \showarticletitle{Building authenticity in social media tools to recruit postsecondary students}.
\newblock \bibinfo{journal}{\emph{Innovative Higher Education}}  \bibinfo{volume}{39} (\bibinfo{year}{2014}), \bibinfo{pages}{333--346}.
\newblock


\bibitem[Schneckenleitner and Thaler(2020)]%
        {schneckenleitner2020use}
\bibfield{author}{\bibinfo{person}{Peter Schneckenleitner} {and} \bibinfo{person}{Johanna Thaler}.} \bibinfo{year}{2020}\natexlab{}.
\newblock \showarticletitle{The Use of Educational Influencers for Communication Activities in the Tertiary Education Sector in Austria}.
\newblock \bibinfo{journal}{\emph{Proceedings of INTCESS}} (\bibinfo{year}{2020}), \bibinfo{pages}{524--531}.
\newblock


\bibitem[Schuth et~al\mbox{.}(2018)]%
        {schuth2018recruiting}
\bibfield{author}{\bibinfo{person}{Marvin Schuth}, \bibinfo{person}{Prisca Brosi}, {and} \bibinfo{person}{Isabell Welpe}.} \bibinfo{year}{2018}\natexlab{}.
\newblock \showarticletitle{Recruiting Women in IT: A Conjoint-Analysis Approach}. In \bibinfo{booktitle}{\emph{51st Hawaii Int. Conf. on System Sciences}}.
\newblock


\bibitem[Shields and Peruta(2019)]%
        {shields2019social}
\bibfield{author}{\bibinfo{person}{Alison~B Shields} {and} \bibinfo{person}{Adam Peruta}.} \bibinfo{year}{2019}\natexlab{}.
\newblock \showarticletitle{Social media and the university decision. Do prospective students really care?}
\newblock \bibinfo{journal}{\emph{Journal of marketing for higher education}} \bibinfo{volume}{29}, \bibinfo{number}{1} (\bibinfo{year}{2019}), \bibinfo{pages}{67--83}.
\newblock


\bibitem[Soylu~Yalcinkaya and Adams(2020)]%
        {soylu2020cultural}
\bibfield{author}{\bibinfo{person}{Nur Soylu~Yalcinkaya} {and} \bibinfo{person}{Glenn Adams}.} \bibinfo{year}{2020}\natexlab{}.
\newblock \showarticletitle{A cultural psychological model of cross-national variation in gender gaps in STEM participation}.
\newblock \bibinfo{journal}{\emph{Personality and Social Psychology Review}} \bibinfo{volume}{24}, \bibinfo{number}{4} (\bibinfo{year}{2020}), \bibinfo{pages}{345--370}.
\newblock


\bibitem[Tahsin et~al\mbox{.}(2022)]%
        {tahsin2022can}
\bibfield{author}{\bibinfo{person}{Noshin Tahsin}, \bibinfo{person}{Nazmus~Sakib Ahmed}, \bibinfo{person}{Moumita Asad}, {and} \bibinfo{person}{Kazi Sakib}.} \bibinfo{year}{2022}\natexlab{}.
\newblock \showarticletitle{Can female underrepresentation in information technology be solved through an awareness-based approach?}. In \bibinfo{booktitle}{\emph{3rd Workshop on Gender Equality, Diversity, and Inclusion in Software Engineering}}. \bibinfo{pages}{1--5}.
\newblock


\bibitem[Tam et~al\mbox{.}(2020)]%
        {tam2020gender}
\bibfield{author}{\bibinfo{person}{Hau-lin Tam}, \bibinfo{person}{Angus Yuk-fung Chan}, {and} \bibinfo{person}{Oscar Long-hin Lai}.} \bibinfo{year}{2020}\natexlab{}.
\newblock \showarticletitle{Gender stereotyping and STEM education: Girls’ empowerment through effective ICT training in Hong Kong}.
\newblock \bibinfo{journal}{\emph{Children and Youth Services Review}}  \bibinfo{volume}{119} (\bibinfo{year}{2020}), \bibinfo{pages}{105624}.
\newblock


\bibitem[Travers et~al\mbox{.}(2023)]%
        {travers2023shooting}
\bibfield{author}{\bibinfo{person}{Marie Travers}, \bibinfo{person}{Jenna Bromell}, \bibinfo{person}{Katie Crowley}, \bibinfo{person}{James~Vincent Patten}, {and} \bibinfo{person}{Ita Richardson}.} \bibinfo{year}{2023}\natexlab{}.
\newblock \showarticletitle{Shooting for the Stars: How a STEM Initiative has Evolved to Address Gender Challenges in Work and Education}. In \bibinfo{booktitle}{\emph{IEEE/ACM 4th Workshop on Gender Equity, Diversity, and Inclusion in Software Engineering (GEICSE)}}. IEEE.
\newblock


\bibitem[Trinkenreich et~al\mbox{.}(2022)]%
        {trinkenreich2022empirical}
\bibfield{author}{\bibinfo{person}{Bianca Trinkenreich}, \bibinfo{person}{Ricardo Britto}, \bibinfo{person}{Marco~A Gerosa}, {and} \bibinfo{person}{Igor Steinmacher}.} \bibinfo{year}{2022}\natexlab{}.
\newblock \showarticletitle{An empirical investigation on the challenges faced by women in the software industry: A case study}. In \bibinfo{booktitle}{\emph{ACM/IEEE 44th Int. Conf. on Software Engineering: Software Engineering in Society}}. \bibinfo{pages}{24--35}.
\newblock


\bibitem[Vooren et~al\mbox{.}(2022)]%
        {vooren2022comparing}
\bibfield{author}{\bibinfo{person}{Melvin Vooren}, \bibinfo{person}{Carla Haelermans}, \bibinfo{person}{Wim Groot}, {and} \bibinfo{person}{Henriette~Maassen van~den Brink}.} \bibinfo{year}{2022}\natexlab{}.
\newblock \showarticletitle{Comparing success of female students to their male counterparts in the STEM fields: an empirical analysis from enrollment until graduation using longitudinal register data}.
\newblock \bibinfo{journal}{\emph{Int. Journal of STEM Education}} \bibinfo{volume}{9}, \bibinfo{number}{1} (\bibinfo{year}{2022}), \bibinfo{pages}{1--17}.
\newblock


\bibitem[Zolduoarrati and Licorish(2021)]%
        {zolduoarrati2021value}
\bibfield{author}{\bibinfo{person}{Elijah Zolduoarrati} {and} \bibinfo{person}{Sherlock~A Licorish}.} \bibinfo{year}{2021}\natexlab{}.
\newblock \showarticletitle{On the value of encouraging gender tolerance and inclusiveness in software engineering communities}.
\newblock \bibinfo{journal}{\emph{Information and Software Technology}}  \bibinfo{volume}{139} (\bibinfo{year}{2021}), \bibinfo{pages}{106667}.
\newblock


\end{thebibliography}


\end{document}